# Good plasmons in a bad metal

Francesco L. Ruta[1,2]*, Yinming Shao[1], Swagata Acharya[3], Anqi Mu[1], Na Hyun Jo[4], Sae Hee Ryu[4], Daria Balatsky[5,6], Dimitar Pashov[7], Brian S.Y. Kim[1,8], Mikhail I. Katsnelson[9], James G. Analytis[6], Eli Rotenberg[4], Andrew J. Millis[1,10], Mark van Schilfgaarde[3], D.N. Basov[1]*

[1]Department of Physics, Columbia University, New York, USA.
[2]Department of Applied Physics and Applied Mathematics, Columbia University, New York, USA.
[3]National Renewable Energy Laboratory, Golden, Colorado, USA
[4]Advanced Light Source, E.O. Lawrence Berkeley National Laboratory, Berkeley, California, USA
[5]Department of Chemistry, University of California, Berkeley, California, USA.
[6]Department of Physics, University of California, Berkeley, California, USA.
[7]Theory and Simulation of Condensed Matter, King's College London, London, UK.
[8]Department of Mechanical Engineering, Columbia University, New York, USA.
[9]Institute for Molecules and Materials, Radboud University, Nijmegen, Netherlands.
[10]Center for Computational Quantum Physics, Flatiron Institute, New York, USA.
*Corresponding authors: fr2441@columbia.edu, db3056@columbia.edu

**Abstract**

Correlated materials may exhibit unusually high resistivity increasing linearly in temperature (1), breaking through the Mott-Ioffe-Regel bound, above which coherent quasiparticles are destroyed (2). The fate of collective charge excitations, or plasmons, in these systems is a subject of debate. Several studies suggest plasmons are overdamped (3-5) while others detect unrenormalized plasmons (6-8). Here, we present direct optical images of low-loss hyperbolic plasmon polaritons (HPPs) in the correlated van der Waals metal $MoOCl_2$ (9). HPPs are plasmon-photon modes that waveguide through extremely anisotropic media (10-15), and are remarkably long-lived in $MoOCl_2$. Many-body theory supported by photoemission results reveals that $MoOCl_2$ is in an orbital-selective and highly incoherent Peierls phase. Different orbitals acquire markedly different bonding-antibonding character (16), producing a highly-anisotropic, isolated Fermi surface (17-18). The Fermi surface is further reconstructed and made partly incoherent by electronic interactions, renormalizing the plasma frequency. HPPs remain long-lived in spite of this, allowing us to uncover previously-unseen imprints of electronic correlations on plasmonic collective modes.

# Main

Van der Waals materials have drawn considerable interest in the last decade as low-loss platforms for polaritons, hybrid light-matter modes that confine electric fields down to the nanoscale (19-20). Surface plasmon polaritons in graphene (21) and hyperbolic phonon polaritons in hexagonal boron nitride (22) and molybdenum trioxide (23-24), in particular, may enable numerous diverse applications including superresolution focusing (25), biosensing (26), negative refraction (27-28) and reflection (29), polarization conversion (30), *et cetera*. Hyperbolic plasmon polaritons (HPPs), which occur in materials with highly anisotropic plasma frequencies, on the other hand, have yet to be widely applied because their host materials have been too lossy (13-14) or even hygroscopic (12,15,31) and non-exfoliatable (32). The realization of practical HPPs will thus open many opportunities for new nanophotonic and quantum devices. In this work, we present evidence of low-loss, steady-state HPPs in the air-stable and exfoliatable van der Waals metal molybdenum oxide dichloride ($MoOCl_2$). We find that HPPs in $MoOCl_2$ can propagate for up to about ten cycles at room temperature – rivaling graphene surface plasmon polaritons (21) – without requiring electrostatic gating or chemical doping. Furthermore, $MoOCl_2$ is found to have an ultrabroad hyperbolic band spanning the near-infrared and visible spectrum, so HPPs can be utilized in frequency ranges that were previously inaccessible to graphene plasmonics.

In addition to technological relevance, low-loss HPPs allow us to probe the intriguing physics of $MoOCl_2$. This compound belongs to the up-and-coming oxychloride family of van der Waals materials (33-34). Oxychlorides are generally characterized by weak interlayer coupling and lattice distortions. Weak interlayer coupling means monolayer and bulk samples share similar properties. Lattice distortions enable exotic anisotropic physics. In the case of $MoOCl_2$, an orbital-selective Peierls distortion produces one-dimensional dimerized Mo chains, splitting Mo-$d_{xy}$ bands away from the Fermi surface (FS) while $d_{xz/yz}$ bands disperse strongly through the Fermi level (16). First-principles studies have suggested that both monolayer and bulk $MoOCl_2$ have large intraband optical anisotropy with interband transitions pushed to higher energies in part due to the Peierls distortion (17-18). Furthermore, $MoOCl_2$ may be regarded as a "bad" metal by some common definitions. $MoOCl_2$ has high carrier density carrier density $2.8 \times 10^{22}$ cm$^{-3}$ but poor room temperature DC conductivity ~1800 $(\Omega \, cm)^{-1}$. $MoOCl_2$ exhibits colossal nonsaturating magnetoresistance and Fermi liquid behavior transitioning to *T*-linear transport above 133 K (9). Similar phenomenology has been noted in high-$T_c$ superconductors (36), some density wave materials (37), and heavy-electron systems (38). In this work, HPPs are used to precisely probe the intraband optical response of $MoOCl_2$ – which is found to differ from predictions by noninteracting theories (16-17). Angle-resolved photoemission spectroscopy (ARPES) interpreted with quasiparticle self-consistent GW (QS*GW*) theory reveals incoherent $d_{xz/yz}$ bands and FS reconstruction by electronic interactions. We may explain the observed HPP dispersion by invoking incoherence in calculations of the optical conductivity of $MoOCl_2$.

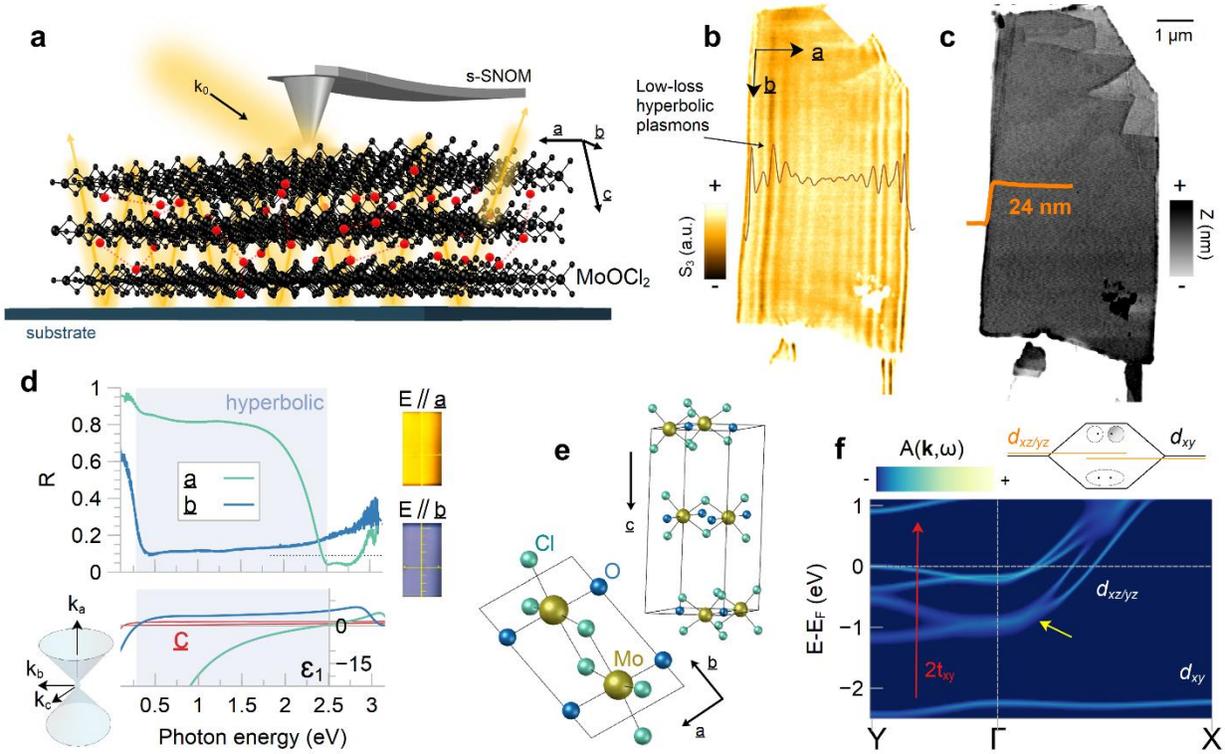

**Figure 1. Hyperbolic plasmon polaritons in molybdenum oxide dichloride: a**, experiment schematic showing the cantilever and tip of a scattering-type scanning near-field optical microscope (s-SNOM) scanning a metallic slab of molybdenum oxide dichloride (MoOCl$_2$) with interacting electrons (red spheres). Free-space light with momentum k$_0$=ω/c scatters off the s-SNOM tip, exciting hyperbolic plasmon polaritons: waveguiding quasiparticles composed of light and electrons. **b,** near-field amplitude (ω=1.312 eV) and, **c**, topography images of a 24-nm-thick MoOCl$_2$ microcrystal. Several unidirectional interference fringes attributable to low-loss hyperbolic plasmon polaritons are observed. **d**, room-temperature polarization-resolved MoOCl$_2$ reflectance spectra reveal large plasma frequency anisotropy (top panel), leading to a wide hyperbolic window where $\varepsilon_1^a \varepsilon_1^b < 0$ and $\varepsilon_1^a \varepsilon_1^c < 0$ (bottom panel) and the isofrequency surface is hyperboloidal (left inset). Hyperbolicity extends into the visible spectrum, attested to by extreme optical anisotropy observable by eye and polarizer (right inset). **e**, bulk MoOCl$_2$ *C2/m* crystal structure (top) and zoom-in of the Peierls-distorted unit cell with Mo dimers along the b-axis (bottom). **f**, theoretical spectral function $A(\mathbf{k},\omega)$ showing incoherent Mo-$d_{xz/yz}$ bands (yellow arrow). Top inset, Peierls distortion causes large bonding-antibonding splitting of $d_{xy}$ band (red arrow, $2t_{xy}$=2.8 eV) while $d_{xz/yz}$ have weaker bonding character.

To access subdiffractional HPPs, we utilize a scattering-type scanning near-field optical microscope (s-SNOM) and a tunable continuous-wave near-infrared laser. Fig. 1a shows a schematic of the experiment, where a metallic s-SNOM tip is illuminated by incident laser light with wavevector $k_0$. The tip launches waveguiding HPPs with wavevector $k$ in a metallic MoOCl$_2$ slab that reflect or transmit at the sample edge, then outcoupling to far-field either via the tip or sample edge. Outcoupled modes will interfere with tip- or edge-scattered light, creating standing wave fringes in real-space scans. Fig. 1b shows a room-temperature s-SNOM image of a thin (Fig. 1c) MoOCl$_2$ microcrystal. Subdiffractional interference fringes attributed to propagating HPPs extend unidirectionally along the a-axis across the entire span of the crystals, highlighting their low damping. A representative line profile with HPP oscillations is shown in brown.

The appearance of HPPs can be understood by studying the dielectric functions of MoOCl$_2$, which can be extracted from Fourier transform infrared (FTIR) spectroscopy measurements (Methods). Bulk MoOCl$_2$ is monoclinic (space group *C2/m*) with lattice parameters a=3.755 Å, b=6.524 Å, c=12.721 Å, and β=104.86° (9), permitting anisotropy between all diagonal (and some off-diagonal) components of the dielectric tensor. Fig. 1d shows FTIR reflectance spectra on bulk MoOCl$_2$ at T=295 K for light polarized along the orthogonal in-plane a- and b-axes of MoOCl$_2$, respectively. The reflectance data show intraband optical responses and the onset of interband transitions above 2.5 eV, which can be mostly explained by an anisotropic Drude-Lorentz model. The model dielectric function $\varepsilon^\alpha = \varepsilon_1^\alpha + i\varepsilon_2^\alpha$ along the α ∈{a, b}-axis is:

$$\varepsilon^\alpha(\omega) = \varepsilon_\infty^\alpha - \frac{(\omega_p^\alpha)^2}{\omega^2 + i\gamma^\alpha \omega} + \sum_n \frac{f_n^\alpha}{(\omega_n^\alpha)^2 - \omega^2 - i\gamma_n^\alpha \omega} \tag{1}$$

where the plasma frequency $\omega_p$ describes oscillations of the electron density, $\varepsilon_\infty$ is the high-frequency permittivity, $f_n$ are the oscillator strengths, $\omega_n$ are the resonance energies, and $\gamma_n$ are the scattering rates. We focus on room temperature results below as measured dielectric functions do not change substantially with temperature (Fig. S1). Anisotropy of $\omega_p$ is the primary driver of hyperbolicity in MoOCl$_2$, causing $\varepsilon_1^a$ to go negative while $\varepsilon_1^b$ and $\varepsilon_1^c$ stay positive (bottom panel) over a broad range $\omega = 0.3$-$2.3$ eV. In this range, MoOCl$_2$ is a quasi-one-dimensional metal: it has a hyperboloidal isofrequency surface (left inset) constraining propagation of extraordinary electromagnetic waves to a conoid centered along the a-axis. Thus, plasmon polaritons may waveguide through the slab and display in-plane hyperbolic wavefronts. The hyperbolic band extends to visible frequencies (dotted line), so even the human eye can detect optical anisotropy with the help of polarizers (right inset). Moreover, we see that dielectric losses in MoOCl$_2$ are small in the hyperbolic band since interband transitions do not appear until $\omega > $ ~2.5 eV (Fig. S1). This rare combination of broad hyperbolic band and weak, spectrally-separated dissipation channels enables low-loss propagation of unidirectional HPPs in MoOCl$_2$.

The large optical anisotropy in MoOCl$_2$ originates in part from an orbital-selective Peierls distortion (16). Mo is in the $d^2$ electronic configuration. In its MoO$_6$ edge-sharing octahedral environment, the $d_{xy}$ orbital is occupied by one electron and $d_{xz/yz}$ orbitals share the other electron. The $d_{xy}$ orbital in such a crystal field environment forms a one-dimensional chain (an Mo-Mo dimer) along the b-axis and a singlet in the b-c plane. Mo-Mo bonds are strongly distorted from dimerization: intra- and inter-dimer bond lengths are 2.7792 Å$^{-1}$ and 3.73648 Å$^{-1}$, respectively (Fig. 1e). As a consequence, the $d_{xy}$ band, which would cross the Fermi level $E_F$ along $\Gamma - Y$ in the undistorted state, picks up strong bonding-antibonding character and splits off from the FS. Meanwhile, intra-dimer hopping between $d_{xz/yz}$ orbitals is too small to form a molecular orbital. $d_{xz/yz}$ bands thus have weaker bonding-antibonding character and still disperse strongly near $E_F$ along $\Gamma - X$ after dimerization, giving MoOCl$_2$ the global metallic behavior observed in transport (9).

Fig. 1f shows the spectral function $A(\mathbf{k}, \omega)$ of MoOCl$_2$ calculated using a self-consistent diagrammatic many-body perturbative theory called QS$GW$ (Methods):

$$A(\mathbf{k}, \omega) = -\frac{1}{\pi} \Im[\omega + \mu - \mathbf{H}_0(\mathbf{k}) - \Sigma(\mathbf{k}, \omega)]^{-1} \qquad (2)$$

where $\mu$ is the chemical potential, $\mathbf{H}_0$ is the one-particle Hamiltonian with the static self-energy $\Sigma_0$, and $\Sigma(\mathbf{k}, \omega)$ is the dynamical part of the self-energy. The spectral function shows flat $d_{xy}$ bands split by $2t_{xy}$ = 2.8 eV (red arrow) while the dispersive $d_{xz/yz}$ bands cross $E_F$. For the most part, QS$GW$ bands qualitatively resemble density functional theory (DFT) eigenvalues (Fig. S2). Importantly, however, in contrast to DFT, QS$GW$ is a many-body approach capable of incorporating the long-range nature of charge scattering. The real part of $\Sigma$ in Equation 2 accounts for energy- and momentum-dependent mass renormalization while the imaginary part contributes to quasiparticle lifetimes and spectral weight redistribution. When including many-body effects beyond the quasiparticle approximation into $\Sigma$, $d_{xz/yz}$ bands display a high degree of incoherence. The outer $d_{xz/yz}$ bands broaden as they move away from $E_F$ (yellow arrow in Fig. 1f). The inner $d_{xz/yz}$ bands also broaden relative to a noninteracting model, and their spectral peaks develop a shoulder (Fig. S3). The quasiparticle weight $Z = (1 - d\text{Re}\Sigma/d\omega)^{-1}$ of $d_{xz/yz}$ bands, representing the overlap between interacting and noninteracting wavefunctions, is ~0.5.

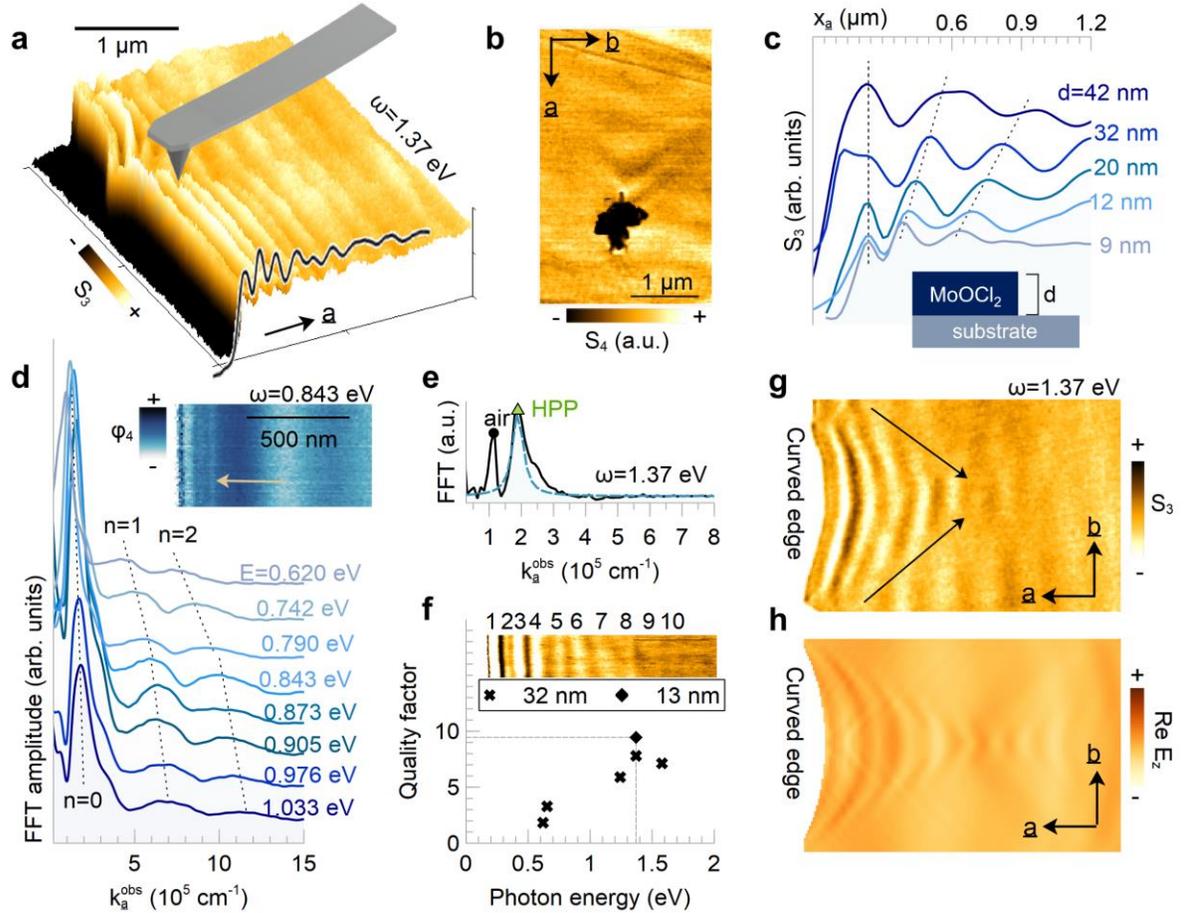

**Figure 2: Near-infrared nano-imaging of hyperbolic plasmon polaritons. a,** room-temperature near-field amplitude image of a 32 nm MoOCl$_2$ microcrystal showing low-loss propagation of hyperbolic plasmon polaritons (HPPs) along the a-axis. Black line is an averaged line profile. **b**, defect-launched hyperbolic wavefronts. **c**, near-field amplitude line profiles from MoOCl$_2$ samples with thicknesses between d=9 nm and 42 nm ($\omega$=1.312 eV). HPP fringes show clear thickness dependence (dotted lines). **d**, fast Fourier transform (FFT) amplitude spectra of complex near-field line profiles with probe energies $\omega$=0.620-1.033 eV. Three dispersive peaks (dotted line) are observed corresponding to the fundamental (n=0) and higher-order (n=1,2) HPP modes. Top inset is a near-field phase image displaying ultra-confined higher-order modes in real-space. **e**, FFT spectrum at $\omega$=1.37 eV showing HPP and air mode peaks. Complex momenta are extracted from Lorentzian fits (dashed blue line). **f**, HPP quality factors $Q = \text{Re } k / \text{Im } k$ reach ~10 at 1.37 eV and have strong energy dependence. Top inset, real-space image of highest-quality sample showing ten fringes. Scale bar is 1 μm. **g**, experimental and, **h**, simulated planar focusing of HPPs at a curved edge.

In Fig. 2, we present additional room-temperature near-field images of $MoOCl_2$ plasmon polaritons inside the hyperbolic band and discuss features uniquely associated with hyperbolicity. In Fig. 2a, the edge of a 32-nm-thick $MoOCl_2$ microcrystal is imaged by s-SNOM under illumination by a near-infrared laser with $\omega$=1.37 eV. The near-field amplitude shows several interference fringes with multiple periodicities which correspond to a superposition of air modes, free-space standing waves between tip and sample, and tip-launched HPPs. On the same sample, we image defect-launched HPPs (Fig. 2b) that show the characteristic hyperbolic wavefronts seen *e.g.* with in-plane hyperbolic phonon polaritons in molybdenum trioxide (23-24). Further, in contrast to evanescent surface plasmons, hyperbolic polaritons behave like waveguide modes with strong thickness dependence and infinite higher-order branches (13,22). In Fig. 2c, we show near-field line profiles from samples of various thicknesses $d$=9-42 nm (Fig. S4). HPP fringe wavelengths shorten, or increase in momentum $k$, with decreasing thickness. This thickness dependence is characteristic of hyperbolic modes (13,22-23,39) – conventional waveguide modes follow the opposite trend (40). Fig. 2d shows fast Fourier transform (FFT) amplitudes of complex near-field profiles at photon energies $\omega$=0.620-1.033 eV. Three dispersive peaks are observed corresponding to fundamental (n=0) and higher-order (n=1,2) HPP modes. Observed n=2 HPPs reach confinement factors $k/k_0 \sim 10$ in our experiments, an order of magnitude tighter than most other reported near-infrared modes (39-40).

In Fig. 2e, we show the FFT amplitude spectrum of a $\omega$=1.37 eV near-field profile. At this higher energy, the air mode and n=0 HPP peaks are well-separated, but higher-order modes are too confined to observe. By fitting Lorentzian parameters to the HPP peak (dashed blue line), we can extract the quality factor $Q = \operatorname{Re} k / \operatorname{Im} k$ (Methods), which expresses the number of propagation cycles of the mode (41). The inset of Fig. 2f shows the real-space image corresponding to Fig. 2e. We count ten fringes and estimate $Q = 9.45$. Quality factors from other samples and laser energies are also plotted in Fig. 2f. Similar quality is measured in the $\omega$=1.305-1.771 eV range, reducing at lower energies where the fringe wavelength also increases significantly. Finally, Figs. 2g and 2h show experimental and simulated (Methods) HPP wavefronts, respectively, at a curved edge demonstrating planar focusing. Similar behavior has been shown for in-plane hyperbolic phonon polaritons and was explained with Huygen's principle, whereby infinite dipoles along the curved edge launch hyperbolic wavefronts interfering into a convex shape (42-43).

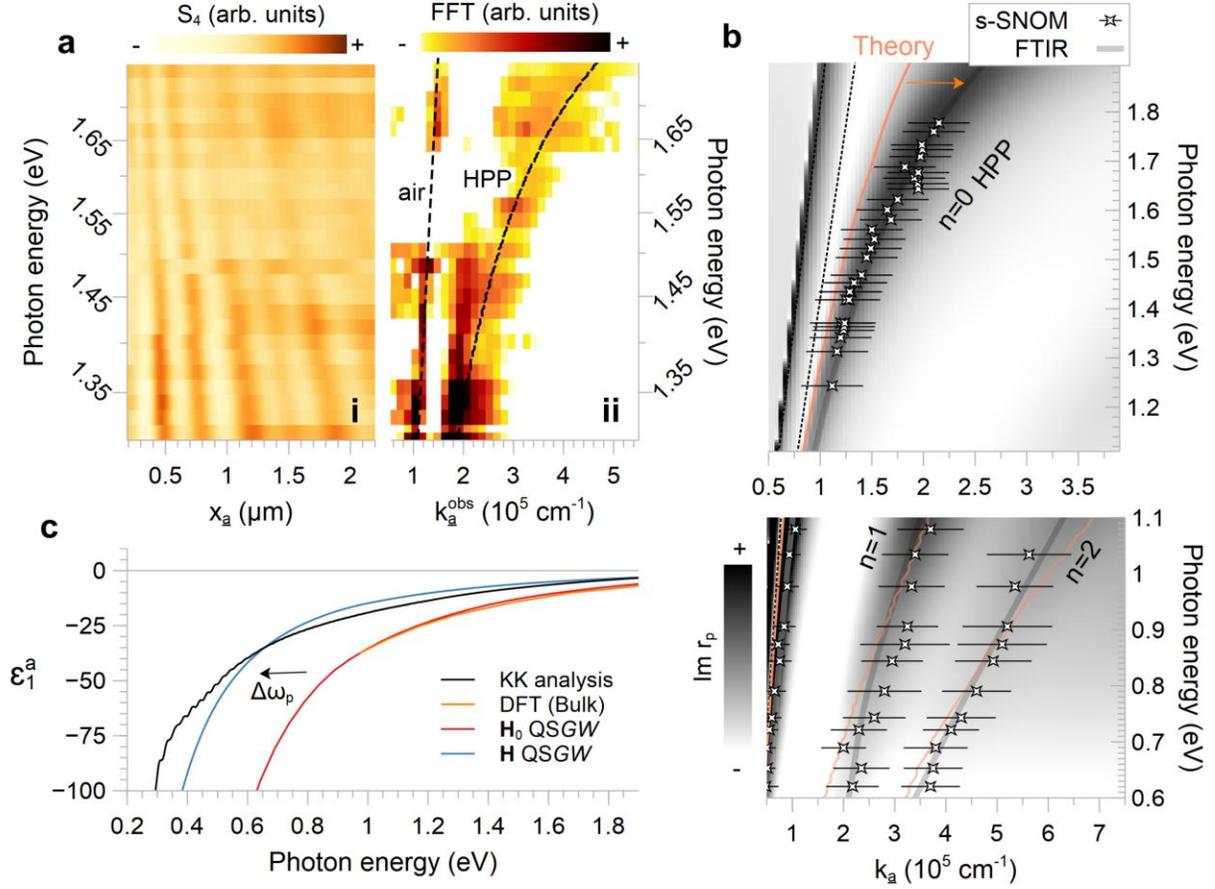

**Figure 3: Electronic correlations renormalize hyperbolic plasmon dispersion. a**, averaged a-axis near-field amplitude line profiles **(i)** and their corresponding fast Fourier transforms (FFT, **ii**) for probe energies $\omega$=1.305-1.771 eV. FFT spectra show the vacuum light cone dispersion (air) and the fundamental (n=0) hyperbolic plasmon polariton dispersion (HPP). Dashed black lines are guides to the eye. **b**, geometry-corrected HPP momenta *k* (error bars are rescaled FFT full-widths at half-maxima) overlaid on dispersions calculated from optical constants extracted from FTIR reflectance fitting and our noninteracting theory (orange line). The colormap is the Im $r_p$ loss function based on the FTIR dielectric function. The observed plasmon dispersion is renormalized to higher momenta compared to calculations from noninteracting theory (orange arrow). Lower energy data with n=1 and n=2 higher-order modes are taken from Fig. 2d and agree with the FTIR dispersion **c**, HPP dispersion renormalization is related to difference in experimental and theoretical plasma frequencies. $\omega_p^a$ obtained from Kramers-Kronig analysis (black) of measured reflectance is ~20% smaller than predicted by density functional theory (DFT, orange) and quasiparticalized *GW* theory ignoring lifetimes and spectral redistribution effects ($H_0$ *QSGW*, red). Inclusion of many-body scattering effects beyond the quasiparticle approximation (**H** *QSGW*, blue) brings $\omega_p^a$ closer to experiment.

Plasmon momenta serve as highly-sensitive, self-referenced probes of intraband electrodynamics (13, 20). We can further constrain the optical parameters of Equation 1 by testing the dielectric function model against the HPP dispersion in addition to FTIR spectroscopy. Plasmon momenta $k$ can be extracted from observed fringe momenta $k^{obs}$ by applying geometrical corrections (Supplementary Note 1). The left panel of Fig. 3a shows averaged line profiles at various probe energies from $\omega =$1.305-1.771 eV. The right panel shows corresponding FFT amplitude profiles displaying two dispersive peaks (black dashed lines). The low-momenta peaks correspond to background air modes at the vacuum light cone. The peaks at higher $k$ are attributed to the fundamental HPP mode. HPPs are observed to the ends of our near-infrared laser range, only disappearing at mid-infrared energies (Fig. S9); confirming the ultra-broadband hyperbolicity of $MoOCl_2$. Also, branching of the dispersion is observed around 1.3 eV (Fig. S8) – the two branches correspond to different paths taken by HPPs to the detector (Supplementary Note 1).

In Fig. 3b, we compute the HPP dispersion from the Im $r_p$ loss function based on the a-axis complex dielectric function extracted from FTIR spectroscopy and theoretical $\varepsilon^c$. Geometry-corrected experimental momenta (white stars) from Figs. 3a and 2d are overlaid on the calculated dispersion, showing reasonable agreement. The HPP dispersion computed from the optics of a noninteracting electronic structure (orange line), on the other hand, systematically underestimates the observed plasmon momentum at higher energies. At lower energies, high-order n=1, 2 modes are mostly sensitive to $\varepsilon^c$ and show good agreement with theoretical $\varepsilon^c$. In the near-field limit and neglecting losses, hyperbolic polaritons in a suspended anisotropic Drude metal with $\varepsilon^c = \varepsilon_\infty = 1$ have dispersion:

$$k_a(\omega) = -\frac{2}{d}\left(1 - \left(\frac{\omega_p^a}{\omega}\right)^2\right)^{-\frac{1}{2}} \operatorname{arctanh}\left(1 - \left(\frac{\omega_p^a}{\omega}\right)^2\right)^{-\frac{1}{2}} \quad (3)$$

where $d$ is the sample thickness. Note that $k_a$ and $\omega_p^a$ are inversely proportional. An HPP dispersion renormalization can thus be attributed to a difference between measured and predicted a-axis plasma frequencies. Indeed, plotting $\varepsilon_1$ in Fig. 3c, the $\sim(\omega_p^a/\omega)^2$ decay of noninteracting theories (both noninteracting QS$GW$ and DFT) implies $\omega_p^a \approx 6.4$ eV. $\omega_p^a = 5.9$ eV may also be obtained directly from the electronic structure (Methods). By contrast, direct Kramers-Kronig analysis (black, Methods) of the FTIR reflectance spectrum, which also fits the HPP momenta as shown in Fig. 3b, implies $\omega_p^a \approx 5$ eV.

To gain insight into possible microscopic mechanisms behind the HPP renormalization, we performed ARPES (Methods). Fermi surfaces in the 3$^{rd}$ and 2$^{nd}$ Brillouin zones and electronic band dispersions measured at a temperature of 11 K are shown in Figs. 4a and 4b, respectively. One closed and one open FS orbit are clearly observed around the Γ point in Fig. 4a (solid white lines) corresponding to the two inner $d_{xz/yz}$ bands. An open orbit was indeed expected from measurements of nonsaturating magnetoresistance (9). Fermi surfaces corresponding to the two outer $d_{xz/yz}$ bands (dotted white lines) are less prominent but may be observed around $k_y = \pm 0.5$ Å$^{-1}$ along with faint pockets around the Y* point (dotted white ellipses). Similar pockets appear in interacting FS calculations (Fig. S10), suggesting that they are related to correlation effects. Exact dimensions of the pockets in theory may be modulated by slightly adjusting $E_F$ within the expected range of error. Further, a faint FS replica is observed along the $k_x$ shifted by ~0.3 Å$^{-1}$ (red arrows), approximately consistent with a predicted charge instability in the static dielectric response at $k_x = 0.27$ Å$^{-1}$ (Methods). Fig. 4c shows a cross-section along $k_y = -0.5$ Å$^{-1}$, where a faint band corresponding to the replica is observed (dashed yellow line). An outer $d_{xz/yz}$ band may have become gapped from a charge density wave (CDW) transition (44-45). Since only one part of the FS, which is anyways incoherent, has become gapped, the CDW transition would not have an observable influence on transport (9). Finally, the -1 eV flat band (black dot) likely developed from photon beam damage, introducing oxygen vacancies that produce localized states (46) along the $\Gamma - X$, reciprocal to the Mo-O bond (Fig. 1e).

Recall that QS$GW$ may include crystal field effects in the self-energy Σ (Equation 2), incorporating the many-body nature of intra-orbital electronic scattering. Many-body effects enhance bonding-antibonding splitting of the $d_{xy}$ band relative to DFT (orange arrow in Fig. 4b) and introduce some degree of incoherence. The measured energy of the bonding band (-1.97 eV) is indeed below the DFT prediction and the high degree of incoherence seen in the QS$GW$ spectral function (Fig. 1f) is also apparent in ARPES, suggesting again that many-body effects may be important. In particular, the outer $d_{xz/yz}$ bands noted earlier are noticeably faded in Figs. 4a and 4b (dotted white lines). At room temperature, the outer $d_{xz/yz}$ bands are hardly observable (Fig. S11). The calculated interacting spectral function at the FS likewise shows a higher intensity of inner bands relative to outer bands (Fig. S10b). In some cases, however, such as the cross-section at $k_y = 0.75$ Å$^{-1}$ at 11 K (Fig. 4d), the outermost $d_{xz/yz}$ band is clearly visible (blue arrows) – suggesting that the observed incoherence is at least in part due to matrix element effects. Nonetheless, ARPES motivates further exploration of interactions and incoherence in understanding the HPP dispersion.

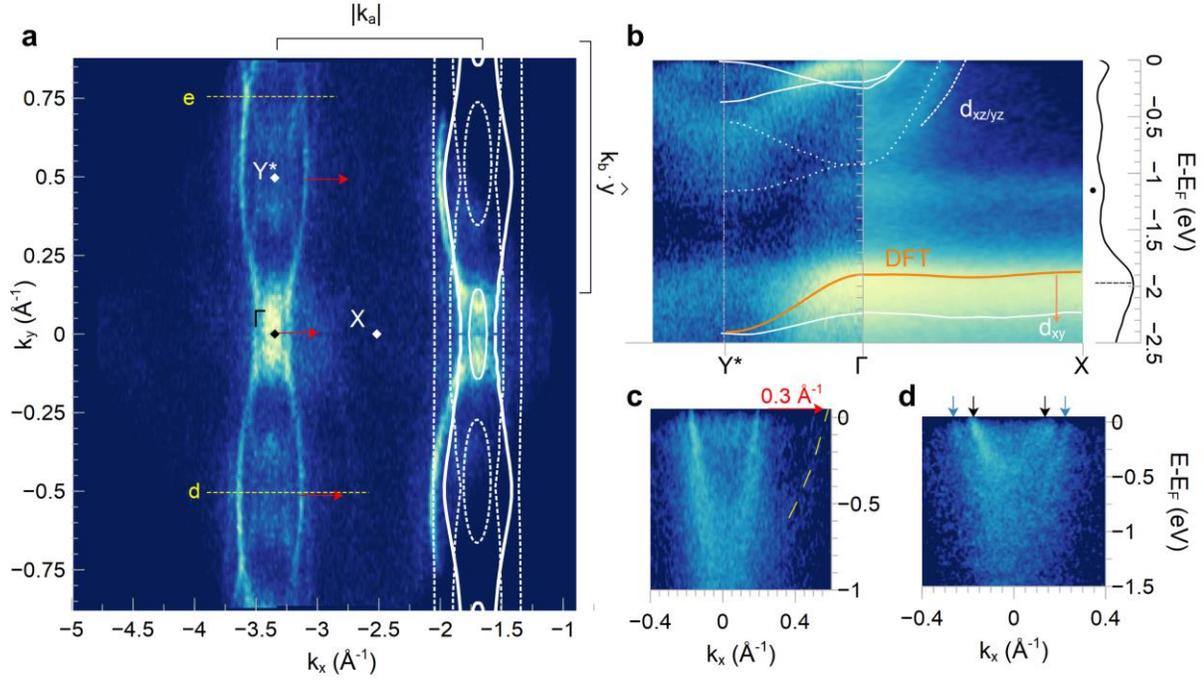

**Figure 4: Angle-resolved photoemission spectroscopy of MoOCl₂ a**, Fermi surfaces measured by angle-resolved photoemission spectroscopy (ARPES) at T=11 K and E=150 eV. White lines are guides to the eye. Prominent open and closed Fermi surface orbits indicated by the solid white lines are referred to as inner $d_{xz/yz}$ bands. Faded bands and Y* point pockets indicated by dashed white lines are outer $d_{xz/yz}$ bands. Faint replica bands are observed shifted by ~0.3 Å⁻¹ (red arrows). **b**, ARPES band dispersions along $\Gamma - X$ and $\Gamma - Y$. QS$GW$ (white) and ARPES are mostly consistent, although the outer $d_{xz/yz}$ (dotted lines) are incoherent. Density functional theory (DFT, orange) underestimates energy of the $d_{xy}$ bonding band (-1.97 eV). Nondispersive -1 eV band (black dot) likely comes from localized O vacancy states. **c**, $k_x$ dispersion at $k_y = -0.5$ Å⁻¹ reveals faint band (yellow) corresponding to replica in **a**. **d**, $k_x$ dispersion at $k_y = 0.75$ Å⁻¹ shows inner (black arrows) and outer (blue arrows) $d_{xz/yz}$ bands of similar intensity.

To account for incoherent bands revealed by ARPES into our optics calculations, we first obtain the real part of the a-axis dynamical optical conductivity $\sigma_1^a$ from the orbital matrix of the QS$GW$ spectral function $A(\mathbf{k}, \omega)$ (47) in our interacting model:

$$\sigma_1^a = \frac{2\pi e^2 \hbar}{V} \sum_{\mathbf{k}} \int d\omega' \, \frac{f(\omega') - f(\omega' + \omega)}{\omega} \times \mathrm{Tr}\{A(\mathbf{k}, \omega' + \omega)v_a(\mathbf{k})A(\mathbf{k}, \omega')v_a(\mathbf{k})\} \quad (4)$$

where $v_a$ are Fermi velocities and $f$ is the equilibrium distribution function. Then, we obtain $\varepsilon_1^a$ by a Kramers-Kronig relation. In Fig. 3c, calculated $\varepsilon_1^a$ with many-body interactions (blue, **H** QS$GW$) is renormalized from the noninteracting picture (red) and represents experimental $\varepsilon_1$ (black) much more faithfully. Similar predictions of reduced plasma frequency in incoherent systems were made by extended dynamical mean-field theory (48). Intuitively, spectral functions, which represent the probability of finding electrons at a given ($\mathbf{k}, \omega$), broaden away from the Fermi level when interactions are introduced (Fig. S3). Electrons are thus less likely to participate in conduction and fewer carriers are effectively available for intraband transitions to quasiparticle states. In a classical picture, incoherence disrupts ordered motion of electrons through scattering and interactions. Electrons will not oscillate in concert: some will be less mobile and averaging over these inefficient oscillations reduces the plasma frequency.

## Conclusion

We have directly observed unidirectional, steady-state HPPs in bulk MoOCl$_2$ microcrystals with record-breaking lifetimes. We confirm the hyperbolic and waveguiding character of plasmons in MoOCl$_2$ with thickness- and energy-dependent measurements. Further, we observed hyperbolic launching and lensing. Next, by obtaining a consistent HPP dispersion between far- and near-field measurements, we extract a precise experimental dielectric function which is renormalized compared to noninteracting *ab initio* calculations. We can reproduce the experimental dielectric function from first principles by including many-body interactions in QS$GW$ theory, which introduces incoherent bands of interacting electrons with reduced plasma frequency. ARPES measurements similarly display incoherent bands and reveal enhanced bonding-antibonding splitting and FS reconstruction induced by electronic interactions. Despite a prevalence of electronic correlations causing incoherent conductivity, HPPs still have weak scattering rates. MoOCl$_2$ is an ideal platform for studying plasmons of correlated electrons, which could shed light on the intriguing physics of "bad" metals. Plus, MoOCl$_2$ is air-stable and exfoliatable, hosting long-lived, ultra-broadband HPPs without electrostatic gating or chemical doping, making it an attractive material for nanophotonics.

## Remarks

We became aware of concurrent work by G. Venturi, *et al.* (49) upon completion of this manuscript.

# Acknowledgments

**Funding:** This work is supported as part of Programmable Quantum Materials, an Energy Frontier Research Center funded by the U.S. Department of Energy (DOE), Office of Science, Basic Energy Sciences (BES), under award no. DE-SC0019443. D.N.B. is a Moore Investigator in Quantum Materials EPIQS no. 9455. M.I.K. is supported by the ERC Synergy Grant, project 854843 FASTCORR (Ultrafast dynamics of correlated electrons in solids). This research used resources of the Advanced Light Source, which is a DOE Office of Science User Facility under contract no. DE-AC02-05CH11231. N.H.J., S.H.R., and E.R. were supported by the Quantum Systems Accelerator (QSA), supported by the U.S. DOE, Office of Science, National Quantum Information Science Research Centers. The Flatiron Institute is a division of the Simons Foundation.

**Author contributions:** F.L.R and Y.S. performed near- and far-field optical measurements. F.L.R. conceived the experiment and did electrodynamics calculations. S.A., D.P. and M.v.S. performed $\text{QS}G\widehat{W}$ calculations. F.L.R. and S.A. analyzed the data. N.H.J., S.H.R., and E.R. performed angle-resolved photoemission spectroscopy. F.L.R. and B.S.Y.K. prepared samples for optical measurements. D.B. and J.G.A. synthesized the $MoOCl_2$ crystals. M.I.K., A.M., and A.J.M. helped to interpret the data. F.L.R. and D.N.B. wrote the manuscript with input from all coauthors.

**Miscellaneous:** We would like to thank Yang Zhang from the University of Tennessee for sharing density functional theory and Wannierization files, which we used to sanity check our own initial noninteracting theory.

**Competing interests:** The authors declare no competing financial interests.

**Data and materials availability:** All data needed to evaluate the conclusions in the study are present in the main text or the supplementary materials.

# Supplementary Materials for

## Good plasmons in a bad metal

*Francesco L. Ruta\*, Yinming Shao, Swagata Acharya, Anqi Mu, Na Hyun Jo, Sae Hee Ryu, Daria Balatsky, Dimitar Pashov, Brian S.Y. Kim, Mikhail I. Katsnelson, James G. Analytis, Eli Rotenberg, Andrew J. Millis, Mark van Schilfgaarde, D.N. Basov\**

Correspond to: *f.ruta@columbia.edu, *db3056@columbia.edu

# Methods

Single crystal growth

Single crystal growth was performed through chemical vapor transport as described by Ref. S1 under argon to prevent oxidation. $MoCl_3$ (99.5% purity) was added to a ~22 cm length quartz ampoule with $MoO_3$ (99.97% purity) in a molar ratio of 2:1. The ampoule was flushed and evacuated with argon three times and sealed under mild vacuum. The ampoule was placed in a horizontal furnace and heated to 400°C at the source and 300°C at the sink at rates of 0.063°C/min and 0.046°C/min, respectively. The sample was held at this gradient for 166.65 hours and then the source and sink were cooled to 300°C and 200°C at a rate of 0.017°C/min and held at those temperatures for 166.65 hours. The source and sink were then cooled to 25°C at rates of 0.028°C/min and 0.018°C/min, respectively. The crystals appeared needle-like with a bronze metallic sheen at the sink of the ampoule. The composition was confirmed with x-ray photoelectron spectroscopy. Also, $MoOCl_2$ crystals purchased from HQ Graphene showed consistent properties with our growth.

Scanning near-field optical microscopy

A Neaspec neaSNOM near-field microscope was used with a continuous-wave tunable laser from M Squared. A Ti:sapphire laser (SolsTiS) was used to cover the wavelength range between 700-1000 nm. By frequency mixing a high-power 532-nm diode laser (Equinox) and the SolsTiS, tunable outputs ranging from 1140-2200 nm were also produced. PtIr-coated NanoWorld Arrow tips nominally with 75 kHz resonance frequencies were used. The signal localized under the apex of the tip is isolated in the backscattered signal by demodulation at the $3^{rd}$-$4^{th}$ tip tapping harmonics. Pseudoheterodyne detection was employed whereby backscattered light is interfered with light from a vibrating mirror and only signal from interference sidebands was collected[S2].

Complex momenta $k$ were extracted from line profiles by taking fast Fourier transforms (FFTs) of near-field profiles and fitting Lorentzian functions to FFT peaks.

$$S(\hat{k}) \propto \left(\left(\hat{k} - \text{Re } k\right)^2 + (\text{Im } k)^2\right)^{-1} \quad (1)$$

Angle-resolved photoemission spectroscopy

Angle-resolved photoemission spectroscopy (ARPES) experiments were conducted at the Beamline 7.0.2 (MAESTRO) at the Advanced Light Source. The microARPES endstation with an Omicron Scienta R4000 electron analyzer was used for data acquisition. Samples were in-situ cleaved by carefully knocking off an alumina post attached to each sample with silver epoxy. Data collection utilized photon energies of 150 eV with a beam size of approximately 15 µm × 15 µm. ARPES measurements were carried out both at T≈11K and room temperature under ultra-high vacuum conditions better than $4 \times 10^{-11}$ Torr.

Fourier transform infrared spectroscopy

Reflectance spectra on $MoOCl_2$ were measured using a Bruker Hyperion 2000 microscope connected to a Bruker Vertex 80V FTIR spectrometer. A tungsten halogen lamp was used as a light source covering a frequency range of 0.5 to ~3.0 eV (near-IR to VIS). A Globar light source was used for the far-infrared (FIR) and mid-infrared (MIR) range from ~0.04 eV to 0.8 eV. Linearly polarized light was focused on the sample using a 15X objective (NA = 0.4) and the aperture size was set to be smaller than the sample dimensions. Reflectance spectra were normalized to an optically thick silver layer (~200 nm) evaporated on the crystal and recorded with a silicon detector (VIS), a liquid-nitrogen-cooled HgCdTe detector (NIR and MIR), and a liquid–helium-cooled silicon bolometer (FIR). Measurements were performed on an as-grown needle of $MoOCl_2$ whose ends were glued to an $SiO_2$/Si substrate with fast-drying silver paint. The surface was cleaved several times with Scotch Magic tape. Temperature-dependent infrared spectroscopy measurements were carried out in a Helium flow cryostat (Oxford MicrostatHe) with a base temperature of 5 K.

Dielectric function fitting was performed with a model consisting of vacuum and $MoOCl_2$ semi-infinite half-spaces. The $MoOCl_2$ dielectric responses along the a and b axes were mostly captured by an anisotropic Drude-Lorentz model (Equation 1 in the main text). Best-fit parameters are listed in Table S1. The c-axis dielectric response was taken from *ab initio* calculations. Fine features of the a-axis dielectric function were independently extracted using a Kramers-Kronig transformation[S3] of the reflectivity measurement with suitable extrapolations. In the low-frequency limit, we use a Hagen-Rubens extrapolation. In the high-frequency limit, we used the calculated x-ray scattering cross-sections followed by a $1/\omega^4$ dependence[S4].

Finite difference time domain simulations

Simulations of the electric fields of hyperbolic plasmon polaritons were performed with the Ansys Lumerical finite difference time domain (FDTD) solver for Maxwell's equations (https://www.lumerical.com/products/fdtd/). We used a point dipole source polarized along the z direction and positioned 20 nm above the top surface of a 54 nm-thick $MoOCl_2$ slab on an $SiO_2$ substrate. Electric field monitors were placed on the top surface and cross-section. The boundary conditions were set to the perfectly matched layer option. To create a curved edge, we place a 100-nm-thick gold cylinder at the edge of the $MoOCl_2$ slab.

*Ab initio* electronic structure and optics

The quasiparticle self-consistent *GW* (QS*GW*) approximation is a self-consistent form of the *GW* approximation[S5]. The noninteracting Hamiltonian $\mathbf{H_0}$ (Slater determinant) that generates **H** is obtained from *GW* in a self-consistent manner. In all simulations, the dimerized phase was used and calculations were performed with periodic boundary conditions along all directions. Single-particle calculations (LDA and energy band calculations with the static part of QS*GW* self-energy $\Sigma_0$) were performed for the 16 atom cell on a 24×24×24 k-mesh while the dynamical self-energy of $\Sigma(\mathbf{k},\omega)$ was constructed using a 6×6×6 k-mesh and $\Sigma_0$ extracted from it. As the off-diagonal components of the quasiparticalized self-energy are calculated, the charge density is updated to be made consistent with the *GW*-derived density, iterating until the root-mean-square change in $\Sigma_0$ reached $10^{-5}$ Ry. Self-consistency eliminates reliance on a density functional and also eliminates any starting point bias. As a result, discrepancies with experiment are more systematic than conventional forms of *GW*[S6-S7]. One of the most severe approximations in this theory is the neglect of electron-hole attraction in the polarizability. QS*GW* has been extended to include ladder diagrams[S8]: for insulators where excitonic effects are important, these diagrams radically reduce the discrepancy between QS*GW* and experiment. For metals and for narrow-gap insulators, on the other hand, excitonic effects are negligible. Thus, QS*GW* (without ladders) generally provides an excellent description of metals[S9] provided correlations are not too strong, as is the case for $MoOCl_2$.

Optics at arbitrary momentum **q** are calculated using the *GW* machinery, generating the polarizability from $\mathbf{H_0}$. This approach includes the Drude term when **q** is large enough that intraband transitions are captured. Thus, it is well suited to find the plasmonic condition at **q**>0 (transitions are not vertical). The dynamical dielectric function $\varepsilon(\omega)$ at low-frequency is largely controlled by the Drude term, which can be parametrized by the plasma frequency $\omega_p$ and the scattering rate $\gamma$ (Equation 1 in the main text). $\omega_p$ was calculated for the QS*GW* noninteracting Hamiltonian $\mathbf{H_0}$[S10]. As expected, it comes out highly anisotropic with values 5.9, 0.96, and 0.32 eV for the a, b, and c components respectively. On the other hand, $\omega_p^a$ deduced from experiment is smaller, as seen from the discrepancy between the QS*GW*/DFT result and measurement in Fig 3c. We traced the cause to the difference between $\mathbf{H_0}$ and **H**. We computed $\sigma_1^a(\omega)$ in the random-phase approximation (RPA) using **H** (Equation 4 in the main text) converged on a 24×24×24 **k**-mesh. Note the exact expression has a vertex; but, for $MoOCl_2$, the RPA is sufficient (as discussed earlier).

We find that including strong frequency dependence of $\Sigma$ is most important for accurately modeling $\varepsilon(\omega)$, bringing the calculated low-frequency $\varepsilon(\omega)$ in good agreement with $\varepsilon(\omega)$ extracted from experiment. N.B. when computing the dynamical self-energy, we do not calculate the full $\Sigma_{ij}(\mathbf{k},\omega)$ tensor to get **H**; but only the diagonal part $\Sigma_{ii}(\mathbf{k},\omega)$. This is because it is costly to compute and, in any case, the diagonal elements are by far the largest components.

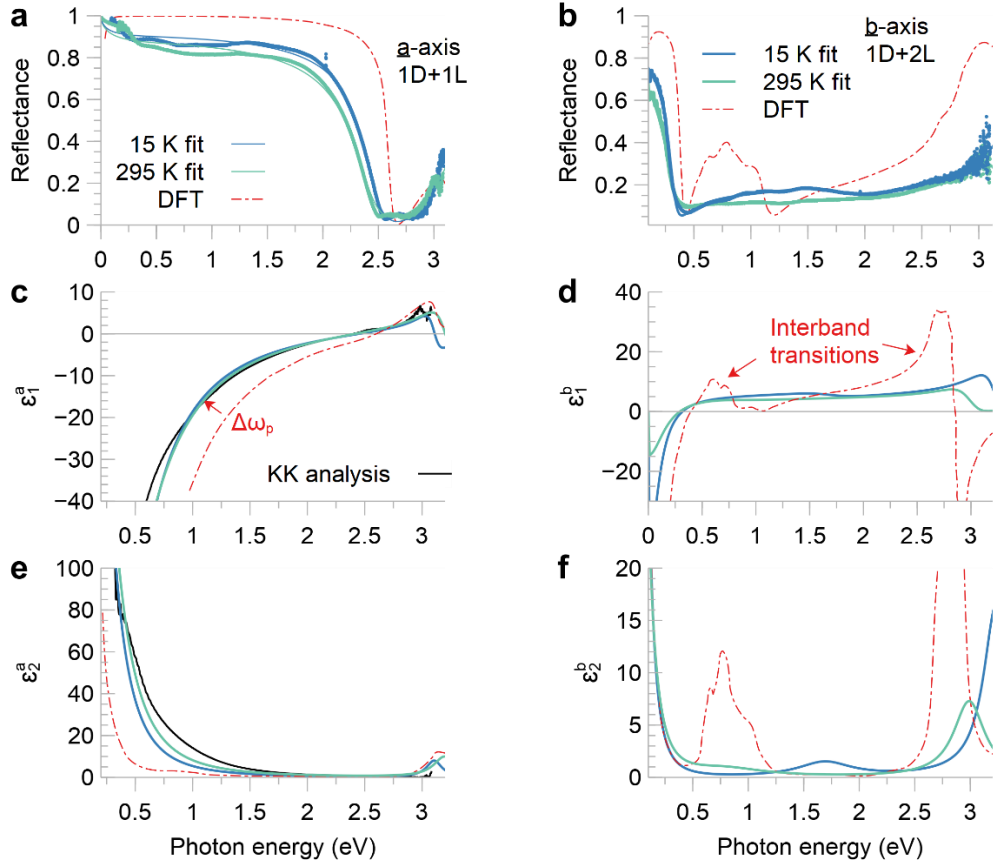

**Figure S1: Far-field infrared spectroscopy and analysis.** Reflectance spectra polarized along the, **a**, a- and, **b**, b-axes of a bulk $MoOCl_2$ microcrystal at T=295 K (cyan) and 15 K (blue). Circles are data and solid lines are Drude-Lorentz fits with best-fit parameters listed in Table S1. Corresponding real (**c-d**) and imaginary (**e-f**) parts of the extracted dielectric functions. Red dash-dotted lines are density functional theory (DFT) calculations from Ref. S11. We observe a lower-energy plasma frequency and weaker interband transitions than predicted by DFT.

**Table S1: Best-fit optical parameters for anisotropic Drude-Lorentz model**

| Crystal axis | Temperature | $\varepsilon_\infty$ (fixed to theory) | $\omega_{0,i}$ (cm$^{-1}$) | $\omega_p \mid \sqrt{f_i}$ (cm$^{-1}$) | $\gamma_i$ (cm$^{-1}$) (*unconstrained) |
|---|---|---|---|---|---|
| a | 295 K | 2.75 | 0 | 41078 | 2875 |
|   |       |      | 25806 | 21692 | 1869.5* |
|   | 15 K  | 2.75 | 0 | 38790 | 2010.1 |
|   |       |      | 25056 | 15989 | 1296.7* |
| b | 295 K | 3.7  | 0 | 6365 | 1424 |
|   |       |      | 7297.8 | 6520.3 | 7618.7 |
|   |       |      | 24153 | 21897 | 2734.5* |
|   | 15 K  | 3.7  | 0 | 6445.6 | 888.78 |
|   |       |      | 13834 | 9537.2 | 4743.5 |
|   |       |      | 26139 | 31446 | 2273.5* |

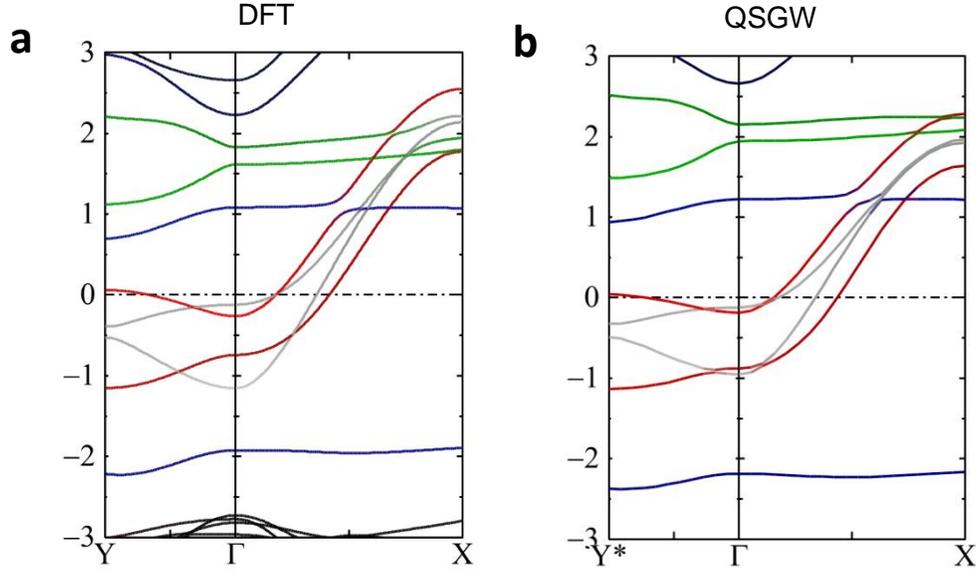

**Figure S2: Comparison of DFT and QS*GW* band structures**. Electronic band structure of MoOCl$_2$ calculated with, **a**, density functional theory (DFT) with the local density approximation (LDA) and, **b**, a QS*GW* theory. The red and grey are d$_{xz,yz}$-like states, blue is d$_{xy}$, and e$_g$-like states are in green.

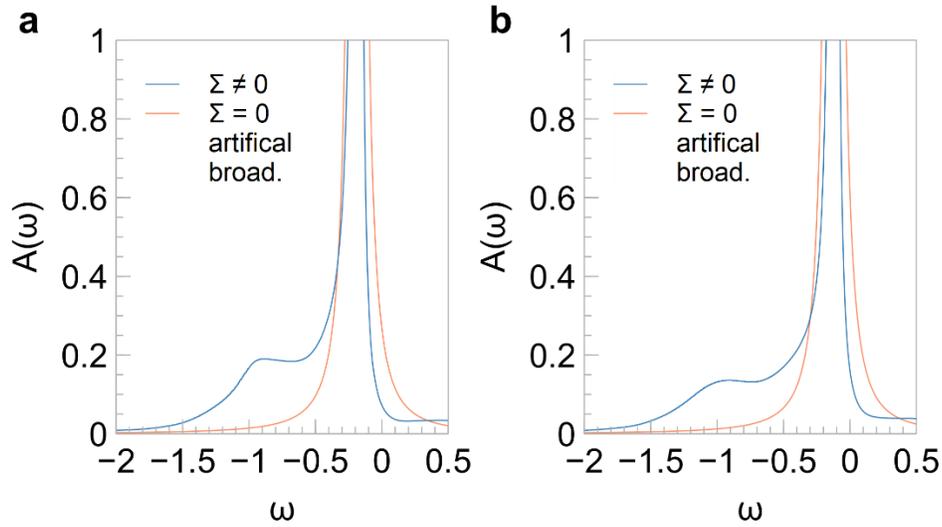

**Figure S3: Spectral functions at Γ point.** Spectral functions corresponding to second-to-innermost (**a**) and innermost (**b**) d$_{xz/yz}$ bands. Both bands develop a low-intensity shoulder around -1 eV when electronic interactions are introduced via the self-energy, moving spectral weight away from the Fermi level.

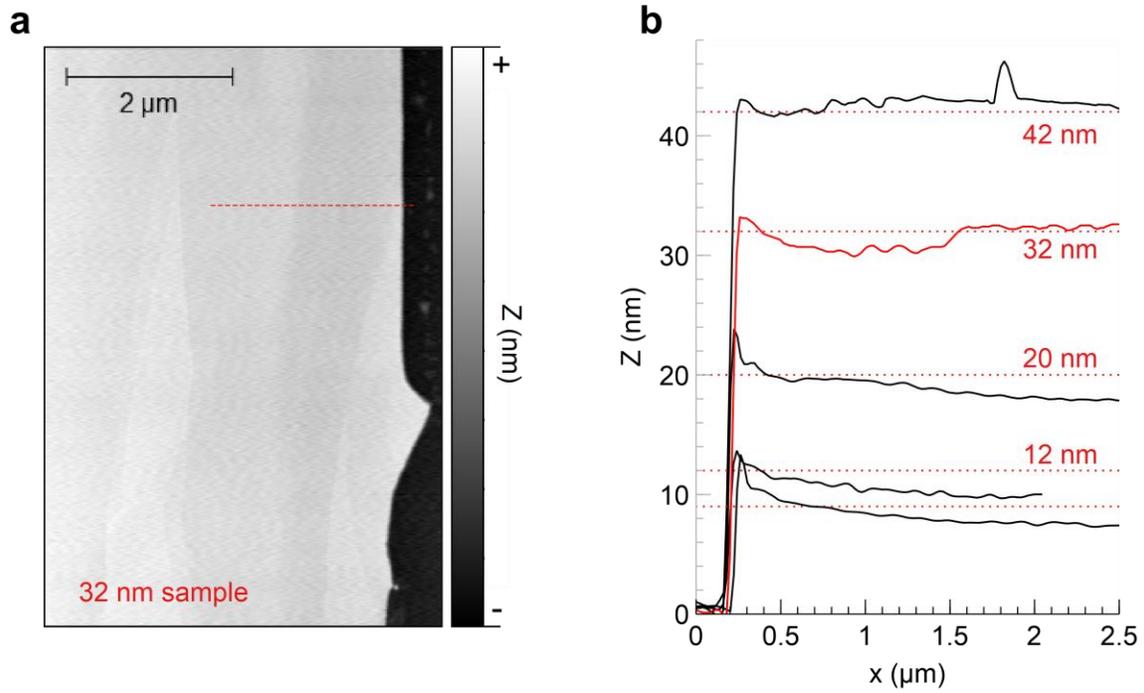

**Figure S4: Atomic force microscopy of MoOCl$_2$ microcrystals. a**, topography image of the 32 nm sample used for dispersion analysis and planar lensing. **b**, topography line profiles from samples used for analyzing thickness dependence of hyperbolic plasmon polaritons in the main text.

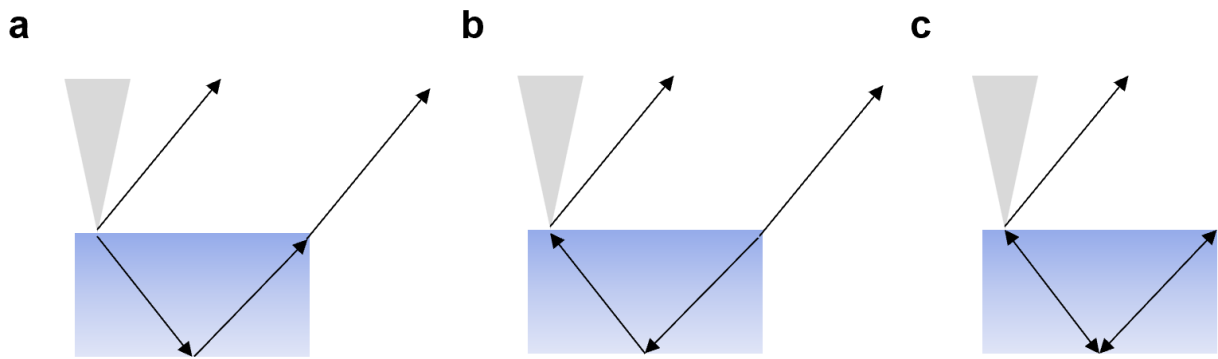

**Figure S5: Propagation paths of polaritons in a slab. a**, edge-transmitted, **b**, edge-launched and, **c**, edge-reflected configurations producing polariton interference fringes in s-SNOM experiments.

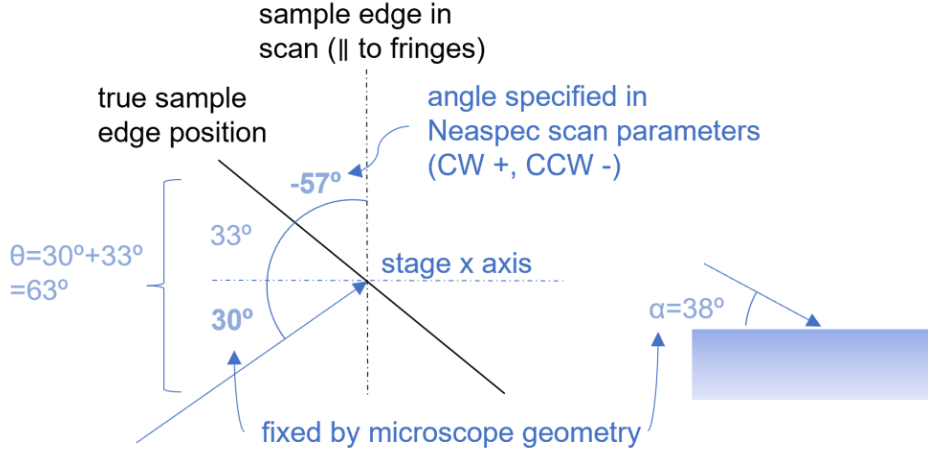

**Figure S6: Sample orientation from Fig. 3 dispersion analysis**. Out-of-plane incident angle $\alpha = 38°$ is fixed by microscope geometry (right). In-plane incident angle $\theta$ is measured relative to the sample edge parallel to analyzed fringes. If the fringes are to the left of the sample edge and the sample edge is oriented vertically in the scan, then $\theta = 30° + 90° + \theta_{scan}$, where 30° is the fixed in-plane incident angle of light relative to the stage x axis and $\theta_{scan}$ is the rotation angle specified in the Neaspec software scan parameters. $\theta_{scan}$ is positive for clockwise and negative for counterclockwise rotations.

**Supplementary Note S1: Geometrical correction to the in-plane wavevector**

Polaritons in a finite slab probed by an s-SNOM tip may propagate along a variety of paths before reaching the detector. Interpretation of standing wave fringes observed in near-field images will differ depending on whether the propagation path is categorized as edge-transmitted, edge-reflected, or edge-launched (Fig. S5). Edge-reflected and edge-transmitted modes are tip-launched and then either reflect or transmit at the edge. They have geometric decay $\sim\sqrt{x}$ associated with spherical wave propagation away from the tip. Edge-reflected modes must complete a round-trip back to the tip before outcoupling to far-field and interfering with light backscattered from the tip. Interfering waves take the same path back to the detector: the fringe wavelength is assumed illumination angle-independent and related to the true momentum $k$ by a constant factor of two[S12]. Edge-transmitted modes and light backscattered from the tip follow different paths back to the detector, on the other hand. Quantitative analysis of edge-transmitted standing wave fringes must account for this path difference by applying an angle-dependent geometrical correction to extracted momenta[S13]:

$$\frac{k}{k^{obs}} = \frac{k - k_0 \sin(\beta - \theta)\cos\alpha}{k\cos\beta} \qquad (2)$$

$$\beta \equiv \arcsin\left(\frac{k_0}{k}\cos\alpha\cos\theta\right) \qquad (3)$$

where $k^{obs}$ is the observed momentum extracted from Fourier analysis, $\alpha = 38°$ is the out-of-plane incident angle[S14], and $\theta$ is the in-plane incident angle relative to the sample edge parallel. Given $k^{obs}$ and $\theta$, one can solve Equations 2 and 3 self-consistently to obtain k. In Fig. S6, we show the orientation of the sample used in the dispersion analysis in Fig. 3 of the main text. If the analyzed fringes are perpendicular to the scan direction, then $\theta = 120° + \theta_{scan}$ where $\theta_{scan}$ is the scan rotation angle specified in the Neaspec NeaSNOM software scan parameters and exported metadata files. The convention used in NeaSNOM is that $\theta_{scan}$ is positive for clockwise and negative for counterclockwise rotations. Finally, edge-launched modes start from the edge and travel to the tip one-way before outcoupling to far-field and interfering with back-scattered light. They usually only appear within $< 1~\mu m$ of the edge[S15]. Edge-launched modes can be thought of as the reverse of the edge-transmitted mode, but without geometric decay. The same geometrical correction used for edge-transmitted modes will apply to edge-launched modes.

Relative to mid-infrared polaritons, the in-plane wavevector of near-infrared polaritons k tends to be closer to the momentum of free-space light $k_0 = \omega/c$. By conservation of momentum, modes launched by the s-SNOM tip thus prefer to transmit through, rather than reflect off of, sample edges and outcouple to the far-field. Edge-transmission tends to dominate for near-infrared modes[S13-S17]. However, plasmon polaritons in MoOCl$_2$ are more confined than previously-observed near-infrared modes and this assumption shoud be carefully revisited.

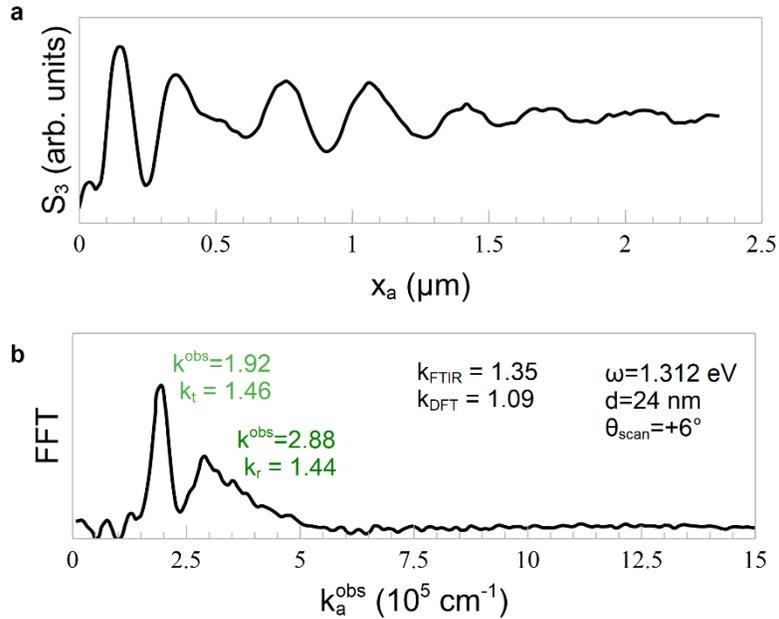

**Figure S7: Coexistence of edge-transmitted and edge-reflected polaritons**. **a**, third-harmonic near-field amplitude line profile from the right edge of Fig. 1b and, **b**, corresponding fast Fourier transform (FFT) showing peaks for both edge-transmitted and edge-reflected polaritons. Correcting observed momenta $k^{obs}$ based on interference paths gives consistent true momenta $k$, which are more consistent with $k_{FTIR}$ implied from the experimentally-extracted dielectric function than $k_{DFT}$, the density functional theory prediction.

In Fig. S8a, we observe branching of the fundamental polariton mode (green triangles). If we apply the edge-transmitted/launched and edge-reflected corrections to the first (light green) and second (dark green) branches, respectively, then their corrected momenta become roughly consistent. This is clearly demonstrated on a 24 nm sample in Fig. S7. If the edge-transmitted/launched and edge-reflected peaks are not well-separated, it becomes unclear which correction should be applied to the composite peak and plasmon momenta may be misestimated. In Fourier spectra where this is the case, we take the average of the two corrections.

In Fig. S8c, the sample is oriented orthogonally ($\theta_{scan} = 22°$) to the sample in Fig. S7a with $\theta_{scan} = -57°$. In Fig. S8c, we find only the edge-transmitted/launched mode. The sample edge in Fig. S8c is oriented parallel to the laser illumination while the edge in Fig. S8a is perpendicular. Dominance of edge-transmitted/launched versus edge-reflected appears to be related both to mode confinement $k/k_0$ and the relative angle between $k$ and $k_0$. The edge-reflected fundamental mode only dominates if $k/k_0 \gtrsim 2.5$ and $k$ and $k_0$ are aligned.

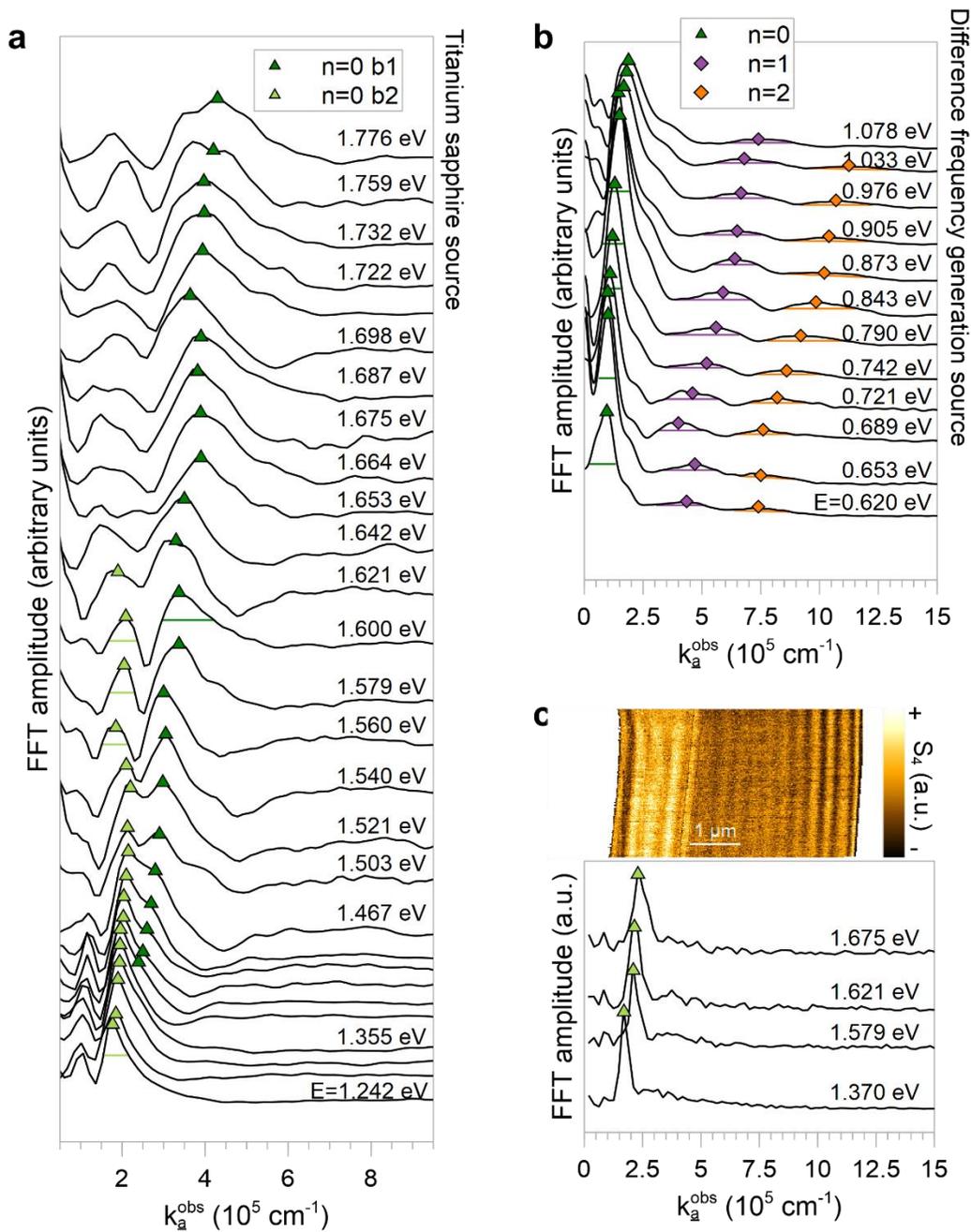

**Figure S8: Fourier analysis of near-field line profiles. a,** fast Fourier transforms (FFTs) of near-field line profiles from a 32-nm-thick MoOCl$_2$ sample at $\theta_{scan} = -57°$ with titanium sapphire laser energies $\omega$=1.242-1.776 eV. Green triangles correspond to points plotted in Fig. 3 of the main text. Two branching hyperbolic plasmon polariton (HPP) peaks are observed at some energies corresponding to coexistence of edge-transmitted and edge-reflected HPPs. **b**, FFT spectra on the same sample with $\omega$=0.620-1.078 eV from a difference frequency generation source. Peaks corresponding to HPPs with mode order up to 2 are indicated. **c**, Fourth-harmonic near-field image (top) and FFT spectra with $\omega$=1.370-1.675 eV (bottom) on a different 26 nm sample with different orientation ($\theta_{scan} = 22°$). Here, only the edge-transmitted modes (light green triangles) are observed.

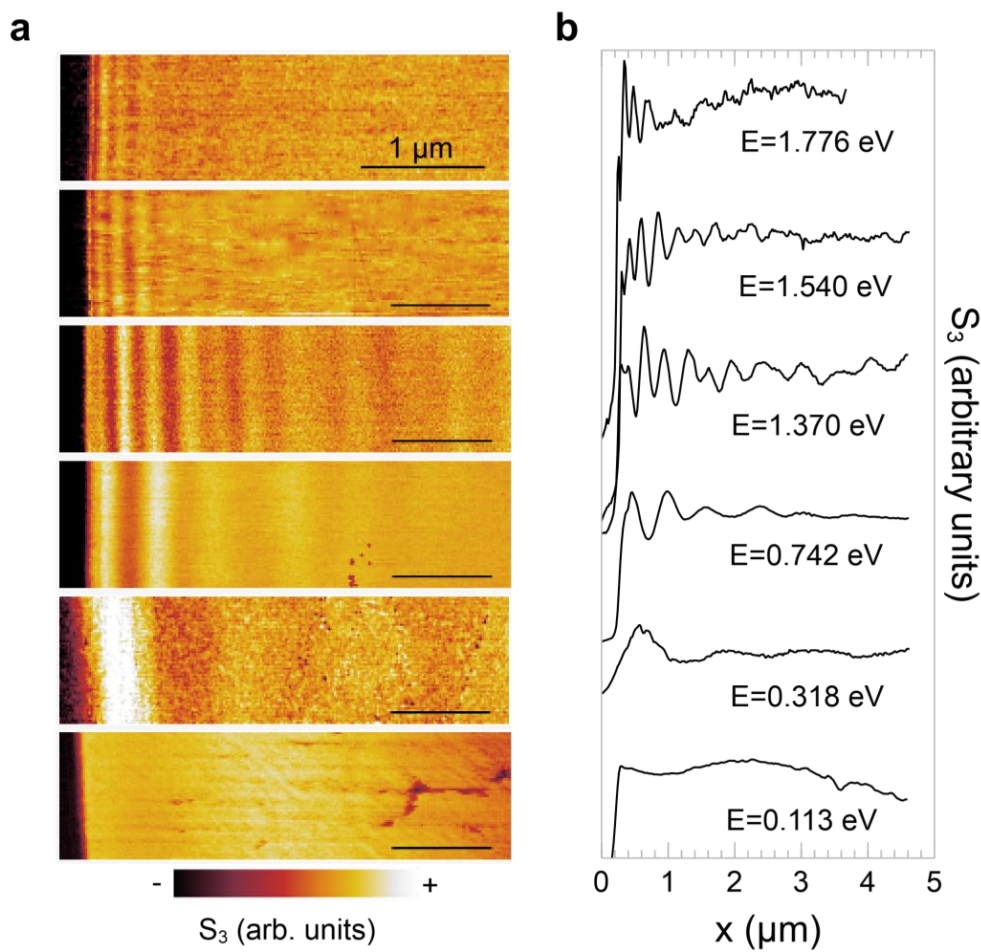

**Figure S9: Visible to mid-infrared nano-imaging of a MoOCl$_2$ microcrystal. a**, third-harmonic near-field amplitude images ranging from visible ($\omega$=1.776 eV, top) to mid-infrared ($\omega$=0.113 eV, bottom) laser frequencies. All scale bars are 1 µm. **b**, corresponding averaged line profiles.

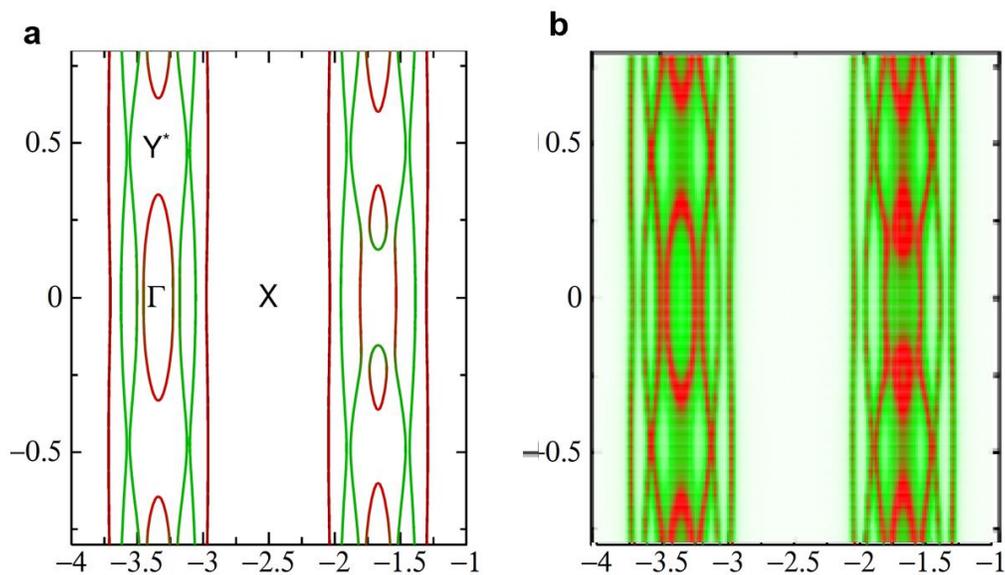

**Figure S10: QS$G\widehat{W}$ Fermi surfaces**. Calculated, **a**, Fermi surface (FS) bands and, **b**, spectral function in the $k_x$-$k_y$ plane. Electronic states in MoOCl$_2$ experience broadening, with spectral weights redistributing over both momentum and energy. After broadening, inner bands show greater intensity than outer bands in the FS spectral function. Furthermore, broadened states lying close to the Fermi level will leave signatures at the FS not observed in band calculations, such as the extra Y* spectral weight observed in both **b** and in experiment.

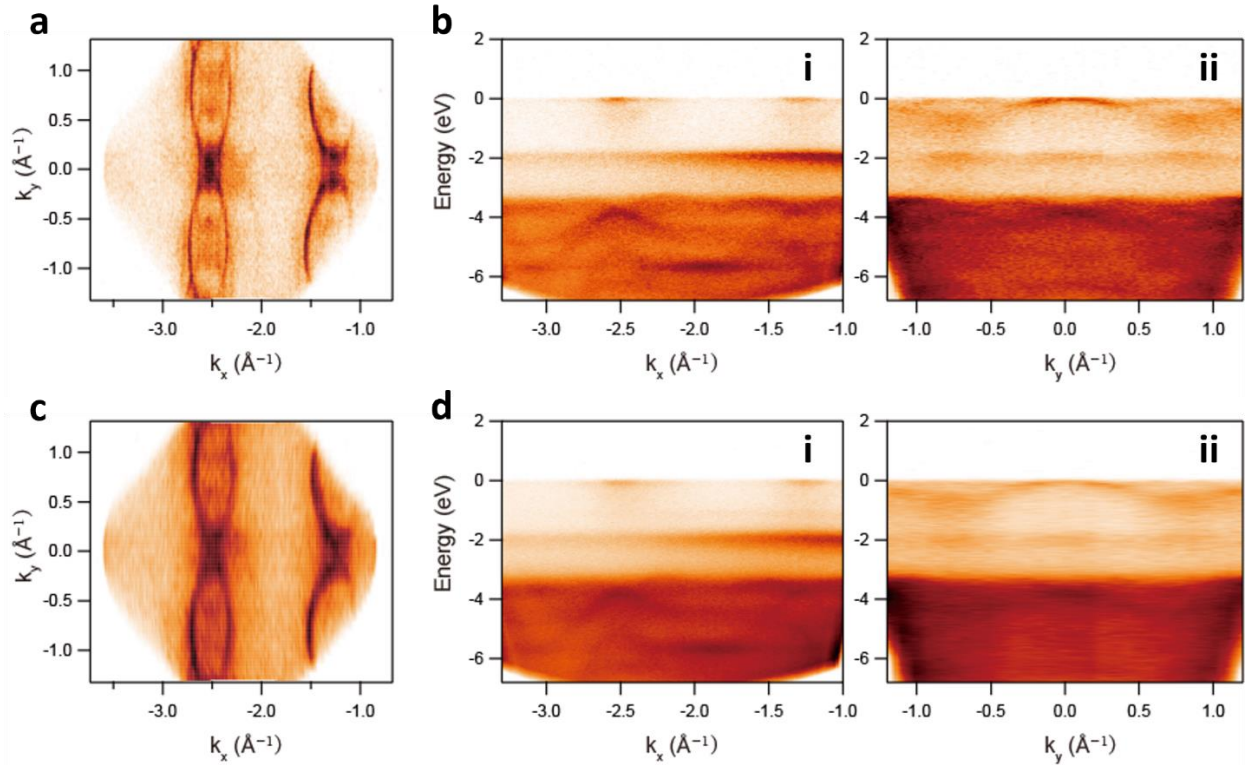

**Figure S11: Temperature-dependent angle-resolved photoemission spectroscopy**. Third and second Brillouin zone Fermi surfaces and $k_x$ (**i**) and $k_y$ (**ii**) dispersions, at, **a-b**, 11 K and, **c-d**, 270 K. Photon energy is 150 eV. No significant temperature dependence of the electronic structure was observed. Faint outer $d_{xz/yz}$ band pockets and Fermi surface replicas are more evident at low temperature, but are nevertheless observable at room temperature.

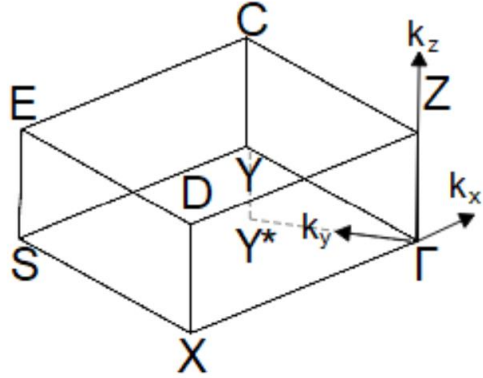

**Figure S12**: **Brillouin zone of dimerized bulk MoOCl$_2$ C2/m lattice.** Reciprocal space coordinates of high symmetry points are specified in Table S3 below.

`

**Table S2: Basis and reciprocal lattice vectors of dimerized bulk MoOCl$_2$ C2/m lattice**

|  | (x, y, z) [Å] | ($k_x$, $k_y$, $k_z$) [Å$^{-1}$] |
|---|---|---|
| **a | $k_a$** | (3.755, 0, 0) | (1.6733, 0, 0) |
| **b | $k_b$** | (0, 6.524, 0) | (0, 0.9631, 0.2555) |
| **c | $k_c$** | (0, −3.262, 12.2956) | (0, 0, 0.5110) |

**Table S3: High symmetry points of dimerized bulk MoOCl$_2$ C2/m lattice**

|  | ($k_a$, $k_b$, $k_c$) | ($k_x$, $k_y$, $k_z$) [Å$^{-1}$] |
|---|---|---|
| Z | (0, 0, 0.5) | (0, 0, 0.2555) |
| Γ | (0, 0, 0) | (0, 0, 0) |
| Y | (0, 0.5, 0) | (0, 0.4815, 0.1278) |
| S | (-0.5, 0.5, 0) | (-0.8366, 0.4815, 0.1278) |
| X | (-0.5, 0, 0) | (-0.8366, 0, 0) |
| D | (-0.5, 0, 0.5) | (-0.8366, 0, 0.2555) |
| E | (-0.5, 0.5, 0.5) | (-0.8366, 0.4815, 0.3833) |
| C | (0, 0.5, 0.5) | (0, 0.4815, 0.3833) |

**Supplementary References**